\documentclass{aa}
\newcommand{\apropto}{\;
  \raise0.3ex\hbox{$\propto$\kern-0.75em\raise-1.1ex\hbox{$\sim$
  }}\;\hskip-2pt }
\usepackage[dvips]{graphicx}
\usepackage[usenames]{color}
\usepackage[T1]{fontenc}
\newcommand{\lta}{\;
  \raise0.3ex\hbox{$<$\kern-0.75em\raise-1.1ex\hbox{$\sim$
  }}\;\hskip-2pt }
\newcommand{\gta}{\;
  \raise0.3ex\hbox{$>$\kern-0.75em\raise-1.1ex\hbox{$\sim$
  }}\;\hskip-2pt }

\begin{document}

\title{Magnetic fields near the peripheries \\of galactic discs
}

   \author{E. Mikhailov\inst{1}
          \and A. Kasparova\inst{2}
           \and D. Moss\inst{3}
          \and R. Beck\inst{4}
          \and D. Sokoloff\inst{1}
           \and A. Zasov\inst{1,2}
}


   \institute{ Department of Physics, Moscow University, 119992 Moscow, Russia
               \and
               Sternberg Astronomical Institute, Moscow M.V. Lomonosov State University, Universitetskij pr., 13, 119992 Moscow, Russia
               \and
               School of Mathematics, University of Manchester, Oxford Road, Manchester, M13 9PL, UK
               \and
               MPI f\"ur Radioastronomie, Auf dem H\"ugel 69, 53121 Bonn, Germany}

   \date{Received 13 Jan 2014 ; accepted 25 June 2014}

\abstract{Magnetic fields are observed beyond the peripheries of
optically detected galactic discs, while numerical models of their
origin and the typical magnitudes are still absent.
Previously, studies of galactic dynamo have avoided considering
the peripheries of galactic discs because of the very limited
(though gradually growing) knowledge about the local properties of
the interstellar medium.} {Here we investigate the possibility
that magnetic fields can be generated in the outskirts of
discs, taking the Milky Way as an example.} {We consider a simple
evolving galactic dynamo model in the ``no-$z$'' formulation,
applicable to peripheral regions of galaxies, for various
assumptions about the radial and vertical profiles of the ionized
gas disc.} {The magnetic field may grow as galaxies evolve, even
in the more remote parts of the galactic disc, out to radii of $
15$ to $30$\,kpc, becoming substantial after  times of about
$10$\,Gyr. This result depends weakly on the adopted distributions
of the half thickness and surface density of the ionized gas
component. The model is robust to changes in the amplitude of the
initial field and the position of its maximum strength. The
magnetic field in the remote parts of the galactic disc could be
generated in situ from a seed field by local dynamo action.
Another possibility is field production in the central regions of
a galaxy, followed by transport to the disc's periphery by the
joint action of the dynamo and turbulent diffusivity.} {Our
results demonstrate the possibilities for the appearance and
strengthening of magnetic fields at the peripheries of disc
galaxies and emphasize the need for observational tests with new
and anticipated radio telescopes (LOFAR, MWA, and SKA).}


\titlerunning{Magnetic fields on the periphery of galactic discs}
\authorrunning{E. Mikhailov et al.}

\maketitle

\section{Introduction}
\label{int} Magnetic fields in spiral galaxies are relatively well
understood observationally and theoretically, at least in several
nearby galaxies (e.g. Donner \& Brandenburg 1990; Subramanian \&
Mestel 1993; Beck \& Wielebinski 2013; Gressel et al. 2013;
Siejkowski et~al. 2014; Brandenburg 2014). However, attention in
such studies has been focussed on galactocentric distances up to
about 10\,kpc, sometimes to 15\,kpc, as in the galaxy M~31 where
magnetic fields are particularly pronounced. Polarized radio
synchrotron emission and its Faraday rotation show that the
regular magnetic fields in this radial range have a spiral form
and, in a number of galaxies, are organized into magnetic arms
that may be situated between material arms, as in e.g. NGC~6946
(Beck 2007; Chamandy et al. 2013; Moss et al. 2013). The typical
strength of the regular magnetic field is several $\mu$G, and the
strengths of magnetic fluctuations are comparable to those of the
regular field.

Regular magnetic fields generated by differential rotation and
mirror asymmetric interstellar turbulence, the mean-field
dynamo\footnote{The mean-field dynamo generates {\em regular}
fields with coherent direction. The magnetic fields traced by
polarized emission can in fact be regular or {\em anisotropic
turbulent}, generated from isotropic turbulent fields by
compression or shear. For the sake of simplicity, the latter field
component is neglected throughout this paper.}, clearly become
weaker at large galactocentric radii, and the polarized radio
emission that traces the regular magnetic field also
decreases towards the galactic periphery. On the other hand, no
sharp radial boundary of the distributions of the relevant
physical quantities is reported. Many papers on the topic (e.g.
Moss et al. 1998) imply that a significant magnetic field may
exist beyond 10\,kpc, because the stellar discs of galaxies (and,
indeed, their gaseous discs) can be traced far beyond the usually
adopted optical radius (usually taken as the radius $R_{25}$
within the isophote of 25th B-magnitude per square arcsec, see
Gentile et al. 2007; Hunter et al. 2011; Holwerda et al. 2012).

We believe that the idea that discs of spiral galaxies can
contain, at substantial galactocentric distances ($r > 10$\,kpc),
a regular magnetic field that is weaker than the field at
$r<10$\,kpc, but still not negligible and possibly dynamically
important, has now become an attractive topic for two reasons.
Firstly, recent observations have revealed that the peripheries of
galactic discs possess pronounced structures. In many cases
spiral-like features are observed in the UV range beyond the main
disc of a galaxy and reveal  continuing star formation, although
with low efficiency (see Bigiel et al. 2010, and references
therein). Another type of outer structure is connected
with the gaseous tidal tails of interacting galaxies -- weak star
formation may also take place there, in spite of the extremely low
mean gas density (Smith et al. 2010).

Although we may assume that the influence of star formation on the
dynamo mechanism may not be very significant (Mikhailov et al.
2012), the presence of young, massive stars provides the ionized
gas layer that is necessary for magnetic field generation.
In addition, the magnetic field plays an active role in
the process of star formation both on large scales (e.g. via the
Parker instability) and small (braking of molecular cloud rotation
or resistance to gravitational compression of a cloud). The
process of star formation in the rarefied gas in the disc
periphery, although not well understood yet,  can therefore be
expected to be different in the presence or absence of magnetic
fields. In some cases non-thermal radio emission is observed that
is clearly connected with tidal features beyond the main bodies of
galaxies, such as M~51 (Fletcher et al. 2011) and NGC~4038/39
(Chy{\.z}y \& Beck 2004). This provides direct evidence that
non-trivial magnetic fields really exist at the peripheries of
discs, playing a significant role in the process of forming of the
outer structures and in star-forming activity in the more rarefied
regions of galaxies. Moreover, it cannot be excluded that these
magnetic fields may influence gas dynamics and, when strong
enough, explain the distorted rotation curves measured from gas
velocities (Ruiz-Granados et al. 2012; Ja{\l}ocha et al. 2012).

Another indication of the presence of extended magnetic
fields in the outer disc comes from the large radial scalelength
of synchrotron emission at about 4\,kpc in many galaxies (Basu \&
Roy 2013), e.g. in NGC~6946 (Beck 2007). In the case of energy
equipartition between cosmic rays and total magnetic fields, the
exponential scalelength of the magnetic field is about four times
greater than that of the synchrotron emission, i.e. about 16\,kpc,
and thus may extend far into intergalactic space (Beck 2007,  see
also e.g. Lacki \& Beck 2013, for discussions about constraints
for equipartition).

Neronov \& Semikoz (2009) and Neronov \& Vovk (2010), from Fermi
observations of blazars in the TeV range, claim that the regular
magnetic field strength in the intergalactic medium is between
$10^{-16}$ and $10^{-9} \mbox{}\mu \mbox{G}$. Mean-field dynamos
cannot excite regular magnetic fields in intergalactic space in
the absence of differential rotation\footnote{The question of
whether a small-scale dynamo can work in intergalactic space
deserves clarification, but is beyond the scope of this paper.}.
Magnetic field transport from galaxies and other objects within
galaxies (for example, active galactic nuclei) is a possibility
for explaining this finding. The presence of a magnetic field that
is relatively weak on galactic scales, but much stronger than
assumed in intergalactic space, would help in such a scenario.

A contemporary observational perspective also looks promising for
studies of galactic magnetic fields on the peripheries of galactic
discs. The point is that the key observations of magnetic fields
at $r \le 10$\,kpc have been performed for wavelengths $\lambda
\le 20$\,cm. Modern radio telescopes are designed to also work at
longer wavelengths of $\lambda \approx 1-10$\,m. In this
wavelength range, observations of regular magnetic fields at
moderate galactocentric distances ($r<10$\,kpc) are strongly
hampered by Faraday depolarization, while the search for weak
magnetic fields at $r>10$\,kpc is a suitable task for these
telescopes. Of course, such investigations require very high
sensitivity.

Previously, dynamo models (e.g. Beck et al. 1996; Moss et al.
2012) did not pay significant attention to the remote parts of
galactic discs, mainly because of the lack of knowledge of the
physical properties of the interstellar medium there. The ideas
available have remained limited until recently, but now sufficient
data have been accumulated to initiate such a discussion. In
principle, the problem could be addressed by various types of
galactic dynamo models. We choose a relatively recent approach
(known as the  no-$z$ model, see below), which allows us to
consider magnetic field evolution in a relatively simple form that
is nonetheless sufficient for our limited knowledge of
hydrodynamics in the remote parts of galactic discs. In the no-$z$
formulation, vertical diffusion is to some extent parameterized,
whereas radial diffusion (which helps by transporting field
outwards in the disc) is explicitly represented.

The aim of this paper is to discuss, for model galaxies similar in
general to the Milky Way, the galactocentric distance out to which
a substantial regular magnetic field might be expected. The paper
is organized as follows. Firstly (Sects.~\ref{WIM} and \ref{rad}),
we summarize  the observational data and their theoretical
consequences for the parameters, relevant to galactic dynamos, of
the shape and hydrodynamics of the remote regions of galactic
discs. In Sect.~\ref{dynamo} we consider a simple galactic dynamo
model applicable to the galactic periphery based on these
governing parameters, in order to obtain magnetic field
distributions in the outer regions. In Sect.~\ref{results} our
results are discussed.

The main finding here is that excitation of magnetic fields in a
radial range out to $20-25$\,kpc is possible, while interstellar
turbulence and differential rotation can in principle maintain a
regular field at even greater galactocentric distances. This
result seems quite robust under changes in models for the
underlying disc structure. In Sect.~5.1 the propagation of
magnetic wave fronts is described: such fronts can also produce
regular magnetic fields in the outer disc regions. In
Sect.~\ref{tests} we discuss possible observational tests for the
presence of magnetic fields in peripheral regions of galactic
discs and give some limitations on the possibilities of detecting
them.

\section{Observations of ionized gas components}
\label{WIM}

For the dynamo mechanism to be efficient, a layer of ionized or
partially ionized gas tied to the magnetic field is needed. Such a
medium exists in real galaxies and consists of two components:
(1)~Cool HI layer (clouds and intercloud medium) partially ionized by
soft cosmic rays and soft X-rays. The ionization of intercloud HI
is about 14\% (see e.g. Berkhuijsen et al. 2006,
Berkhuijsen~\&~M\"uller 2008). A neutral component is tightly
coupled to the magnetic field due to collisions between neutral
and ionized atoms. (2)~It also contains a diffuse warm
ionized medium (WIM) that is mostly identified from optical
recombination lines and pulsar dispersion measures (DMs). Its
ionization is supported by ionizing photons that escape from the
sites of star formation, and partially also from pre-white dwarfs
(see Ferri{\`e}re 2001 and Haffner et al. 2009 for reviews).

Observations give direct information concerning only the WIM
parameters in the vicinity of Sun ($R_{\odot} \approx8.5$\,kpc),
based on pulsar DMs and H${\alpha}$ emission. The averaged volume
density of the WIM and its half thickness obtained by different
authors lie in the ranges $0.015-0.03$ cm$^{-3}$ and
$700-1800$\,pc, respectively (see Berkhuijsen et al. 2006;
Gaensler et al. 2008; Schnitzeler 2012 and references therein).
The thickness of the WIM layer exceeds what is expected for a
layer in hydrostatic equilibrium with the observed temperature,
presumably because of random (turbulent) gas motions. The latter
may be characterized by a turbulent velocity (which we identify in
the framework of this paper with the speed of sound) that is
higher than the usual speed of sound in a thermal plasma.  The
vertical profile of the WIM found from the pulsar DMs is rather
shallow, which may partially be due to the increase in its filling
factor along the coordinate perpendicular to the disc, up to the
height $|z| \approx 900$\,pc. For larger $|z|$ the mean value of
$\rm{DM}_{p}=\rm{DM}\cdot \sin |b|$ ceases to increase (Cordes \&
Lazio 2003; Berkhuijsen et al. 2006), which provides evidence of a
sharp decrease in electron number density.

Unfortunately, the half thickness of the ionized medium at
different galactocentric radii $r$ in the Galaxy is poorly known,
although some data suggest that its distribution is shallower than
the radial density profile of the stellar disc (Cordes \& Lazio
2002; Cordes \& Lazio 2003). In spite of the divergence of the
data, a robust estimate of DM$_{\rm p}$ perpendicular to the disc
plane is given by different authors as $\rm{DM}_{\rm p}
\approx20-26$~cm$^{-3}$\,pc (Cordes \& Lazio 2002; Schnitzeler
2012; Gaensler et al. 2008), which corresponds to a column density
of the WIM for the solar vicinity of $\Sigma_{{\rm
i},0}=2$\,DM$_{\rm p}$\,m$_{\rm p} \approx 1$~M$_\odot$/pc$^2$,
where m$_{\rm p}$ is the proton mass.

\section{The parameter distributions}
\label{rad}

To study galactic dynamos in the no-$z$ approximation, we need to
know the radial distribution of the half thickness of the dynamo
layer, its mean volume density, and the angular velocity of the
rotating gas, $\Omega(r)=V_{\rm rot}(r)/r$. The magnetic field
growth is described by the dynamo number, which can be
parameterized as $D=(3 \, h_{\rm dyn} \, \Omega/v)^{2},$ where
$h_{\rm dyn}$ is the dynamo scaleheight (i.e. an estimate of the
scale over which the magnetic field is expected to vary
perpendicular to the disc), $\Omega$ is the angular velocity, and
$v$ is the velocity of turbulent motions. We emphasize that
$h_{\rm dyn}$ can only be an order-of-magnitude estimate, and it
is probably not realistic to identify it precisely with any
specific physical scale. This approach has been verified as
satisfactory in a number of cases, and we show that our
conclusions about the existence of a dynamo-generated field beyond
the normal boundaries of galaxies are not sensitively
dependent on this choice. The growth of magnetic field from
dynamo action is a threshold process: the field can grow if
$D>D_{\rm cr} \approx 7$ (Phillips 2001; Arshakian et al. 2009).

Observations give the rotation curve $V_{\rm rot}(r)$ of the
Galaxy up to $14-16$\,kpc from the galactic centre (Brand \& Blitz
1993; Vall\'ee 1994; Bovy et al. 2012). For our purposes it is
sufficient to adopt a flat rotation curve in the range $5-30$\,kpc
with a plateau at 220~km~s$^{-1}$. We also tested a more detailed
shape of the rotation curve; however, this does not significantly
affect the results.

Taking the uncertainties in the data into account, we consider two
simple models with very different, however feasible, parameters
for the layer of partially ionized gas (WIM+HI). As the basic
starting point, we assumed that this layer is not too thin,
otherwise a magnetic field could not be generated (Phillips 2001;
Arshakian et al. 2009; Moss \& Sokoloff 2011). Also, the observed
half thickness of the layer of partially ionized warm HI in the
solar vicinity is about $400$\,pc (Cox 2005). The radial profile
of the half thickness of this layer in the outer parts of the
Galaxy is also little known, so we consider two different cases: a
layer with constant half thickness and a layer that flares at
large radial distances. Both models are normalized by the column
density of ionized medium in the solar vicinity of about
$1$~M$_\odot$/pc$^2$, derived from pulsar observations (see the
previous section). The radial profile of the column density
$\Sigma_{\rm i} (r)$ in both models is assumed to be proportional
to the HI profile $\Sigma_{\rm HI}(r)$. Following Kalberla et al.
(2007), we represent the azimuthally averaged HI surface density
by $\Sigma_{\rm HI}(r) = 10$~M$_\odot$/pc$^2$ for $r<12.5$\,kpc
and $\Sigma_{\rm HI}(r) = s_0\cdot e^{-(r-R_{\odot})/R_{s}}$ for
$12.5< r<30$\,kpc, where $s_0=30$~M$_\odot$/pc$^2$ and
$R_{s}=3.75$\,kpc.

\underline{Model~1.} The dynamo layer has a constant scaleheight
$h_{\rm dyn} = 1000$\,pc. In this case the velocity dispersion of
ionized gas should decrease from 40 to 9~km\ s$^{-1} $ in the
range $5-25$\,kpc in order to agree with the gravitational
potential of the disc. This parametrization of the scaleheight is
close to the half thickness of the warm ionized gas layer (see
previous section).

\underline{Model~2a.} The dynamo scaleheight is $h_{\rm dyn} =
400$\,pc up to $18$\,kpc. This scaleheight is close to the half
thickness of the layer of partially ionized warm HI. We assume
here that it remains constant even at large distances from the
galactic centre. Since the partially ionized layer cannot be
thinner than the HI layer, we assumed that for $r>18$\,kpc $h_{\rm
dyn}$ is close to $h_{\rm HI}$, which grows exponentially with $r$
as is observed in our Galaxy: $h_{\rm HI} = h_0\cdot
e^{(R-R_{\odot})/R_0}$, where $h_0=0.15$\,kpc and $R_0=9.8$\,kpc
(Kalberla et al. 2007). The turbulent velocity of gas in this
model is taken as 10\,km\ s$^{-1} $ which is usually valid for the
peripheries of the discs of spiral galaxies (Walter et al. 2008).
The half thickness for $10 \mbox{ kpc}<r<18 \mbox{ kpc}$ in this
model is small, so that the dynamo number is not very large in
this region. However, the magnetic field grows in the outer parts.
We can conclude that for more realistic cases, the magnetic field
strength will be at least as large as in this model.

\underline{Model~2b.} This model is similar to Model~2a. The
difference is that at large distances ($r>12.5$\,kpc), $h_{\rm
dyn}$ was taken not from the Kalberla et~al.~(2007) approximation,
but from a self-consistent equilibrium model of the gas layer,
situated in the gravitational potential of the galactic disc and
 dark halo (see Kasparova \& Zasov~2008 for details).  In this model $h_{\rm dyn}$
at large radii grows more steeply  with $r$  than in Model~2a.
The advantage of this model is that in principle it could be applied to
other galaxies where, unlike the case of our Galaxy,  an estimate for
the thickness of the gas layer is not known directly from observations.

The early galactic dynamo models (e.g. Ruzmaikin et al. 1985)
preferred $v=10\mbox{ km s}^{-1}$ and relatively thin discs with
$h_{\rm dyn} = 400$\,pc. However, after the structure of
interstellar gas became clearer, galactic dynamo models with
thicker discs were considered (e.g. Poezd et al. 1993), and the
choice of the disc thickness depends on the rms turbulent velocity
adopted. Lower turbulent velocities were considered, retaining the
other dynamo parameters, this would assist dynamo action because
the dynamo number is inversely proportional to $v^2$.

\section{A simple dynamo model for  the remote parts of galactic discs}
\label{dynamo}

\begin{figure*}
\begin{center}
\includegraphics[width=12cm]{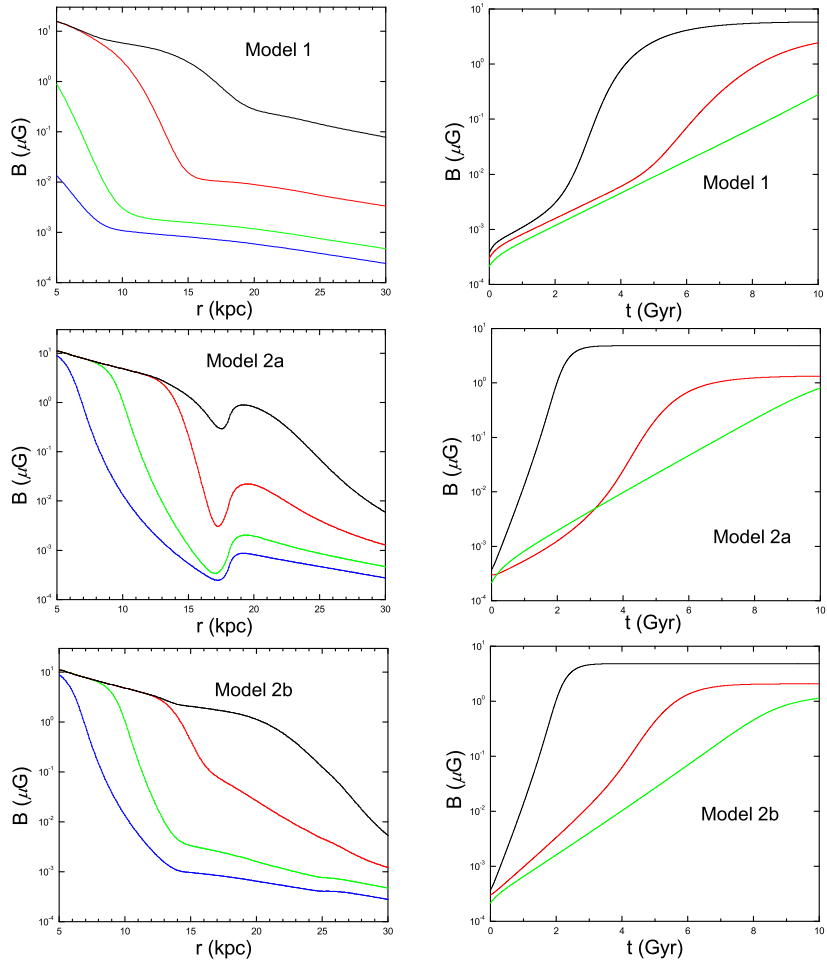}
\end{center}
\caption{Left: radial profile of $B$ for three different profiles
of the dynamo scaleheight (see text). The curves from bottom to
top in each panel indicate magnetic fields at times $t=1,\ 2,\ 5,
\mbox{ and } 10 \mbox{ Gyr}$. Right: growth of the large-scale
magnetic field at $r=$10, 15, and 20\,kpc (from top to bottom in
each panel).} \label{fig2}
\end{figure*}

To model the magnetic field evolution in the outer parts of
galactic discs, we used a simple model in the no-$z$ formulation
(e.g. Subramanian \& Mestel 1993; Moss 1995 and subsequent
papers), taking the $\alpha \omega$ approximation. The model was
formulated  for the field components parallel to the disc plane
($B_r$ and $B_\phi$) with the implicit understanding that the
component perpendicular to this plane (i.e. in the $z$-direction)
is given by the solenoidality condition and that the field has
even parity with respect to the disc plane\footnote{We have
checked a posteriori that $\partial B_{z} / \partial z \approx
10^{-4}$ $\mu$G/{pc} even in the inner parts (in the outer parts
it is less). Assuming $|\partial B_z/\partial z|\approx |B_z/h|$
and $h \approx 400$\,pc, we typically obtain $B_z \la
0.04$~$\mu$G, which is consistent with the limits of the vertical
field strength derived from observations of Faraday rotation
measures (Mao et al. 2010).}. The field components parallel to the
plane can be considered as mid-plane values, or as a form of
vertical average through the disc (see e.g. Moss 1995). We assume
that the large-scale magnetic field is axisymmetric.

The no-$z$ dynamo equations in cylindrical polar coordinates ($r,
\varphi, z$) can be written as
\begin{equation}
\frac{\partial B_{r}}{\partial \tau}=-\frac{\alpha
B_{\varphi}}{h_{\rm dyn}} +\eta \left\{ -\frac{\pi^{2} B_{r}}{4h_{\rm dyn}^2}+\frac{\partial}{\partial
r} \left(\frac{\partial}{r \partial r}(rB_{r})\right) \right\};
\label{eqBr}
\end{equation}
\begin{equation}
\frac{\partial B_{\varphi}}{\partial \tau}=r \frac{\partial \Omega}{\partial r} B_{r} +\eta \left\{
-\frac{\pi^{2} B_{\varphi}}{4h_{\rm dyn}^2}+\frac{\partial}{\partial r}
\left(\frac{\partial}{r \partial r}(rB_{\varphi})\right) \right\},
\label{eqBphi}
\end{equation}
where $\alpha$ characterizes turbulent motions, $\Omega$ is the
angular velocity, $\eta = l \, v / 3$ the turbulent diffusivity,
$l$ the length scale of turbulence (taken to be constant
throughout the galaxy), and $v$ the rms turbulent velocity
(Arshakian et al. 2009). The $z$~component does not appear
explicitly, and the equations have been calibrated by factors of
${\pi^{2}}/{4}$ in the vertical diffusion terms (Phillips 2001).
We use the system (\ref{eqBr}), (\ref{eqBphi}) in dimensional form
to aid physical interpretation of our results. Distances are
measured in kpc, times in Gyr, and magnetic field strengths in
$\mu \mbox{G}$.

\begin{figure}
\begin{center}
\includegraphics[width=8cm]{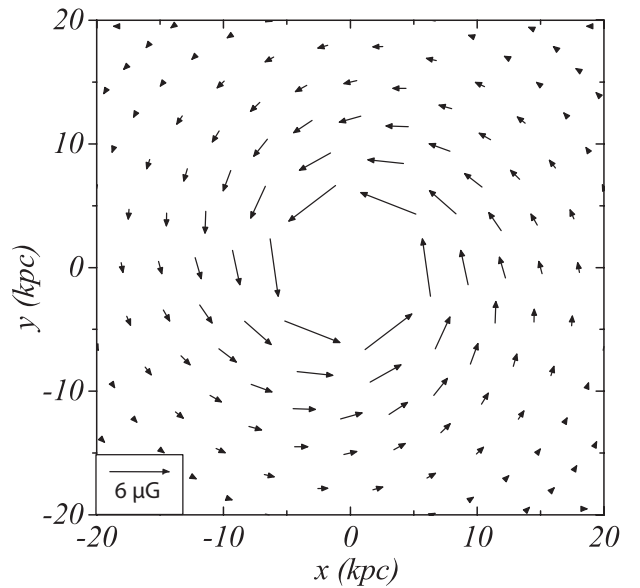}
\end{center}
\caption{Typical vector plot of the large-scale magnetic field (which is almost azimuthal).}
\label{fig3}
\end{figure}

\begin{figure}
\begin{center}
\includegraphics[width=7cm]{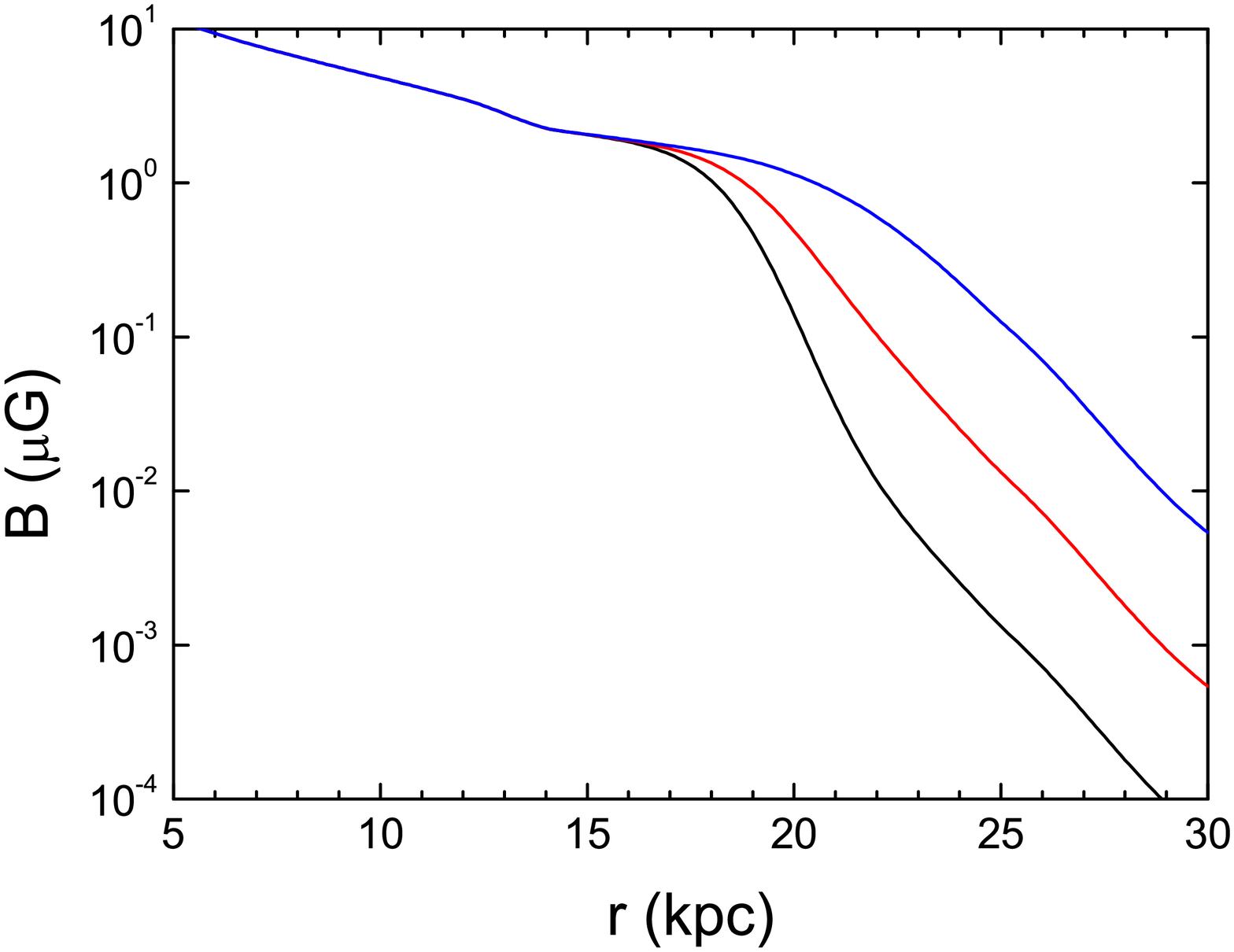}
\includegraphics[width=7cm]{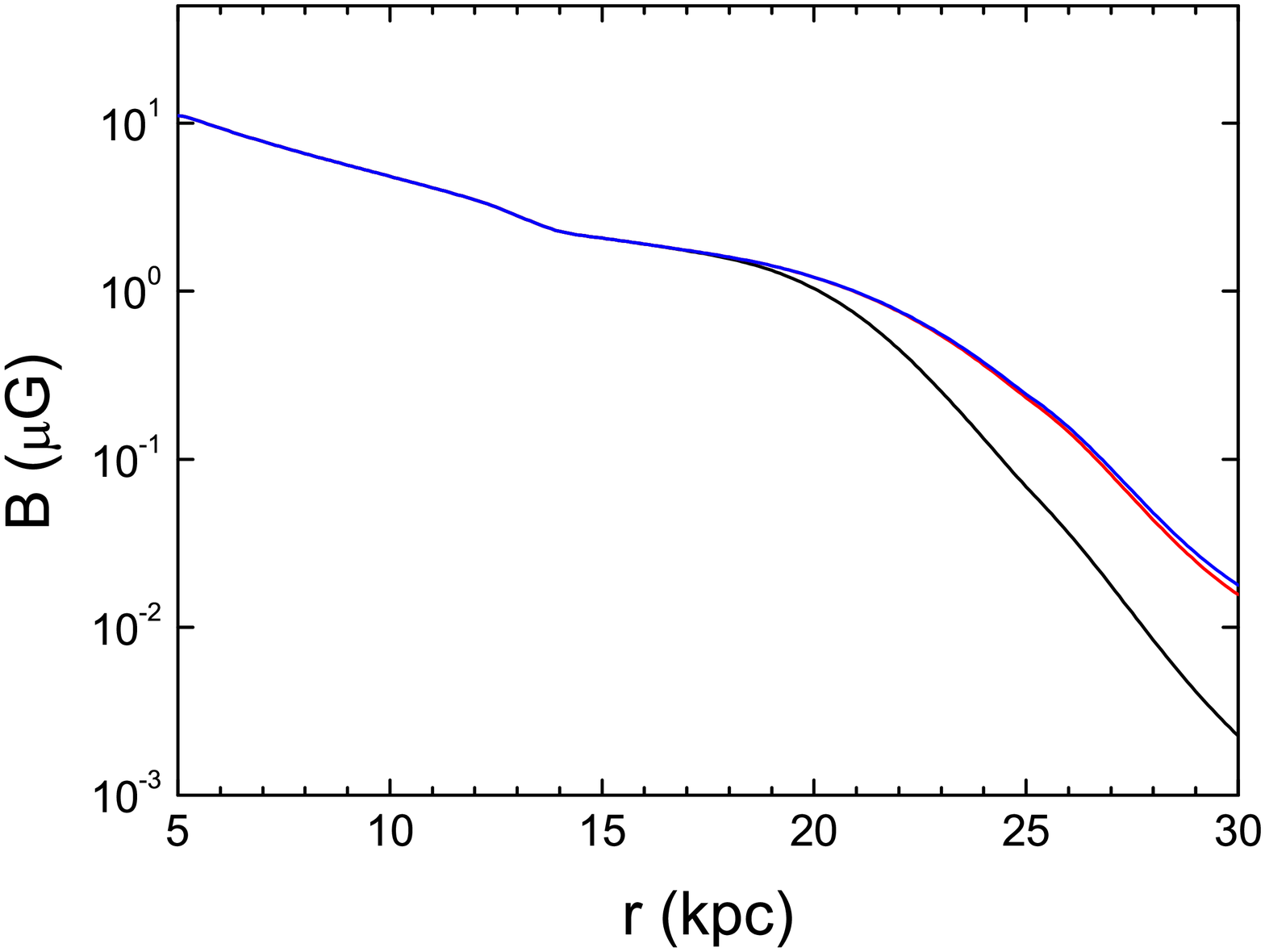}
\includegraphics[width=7cm]{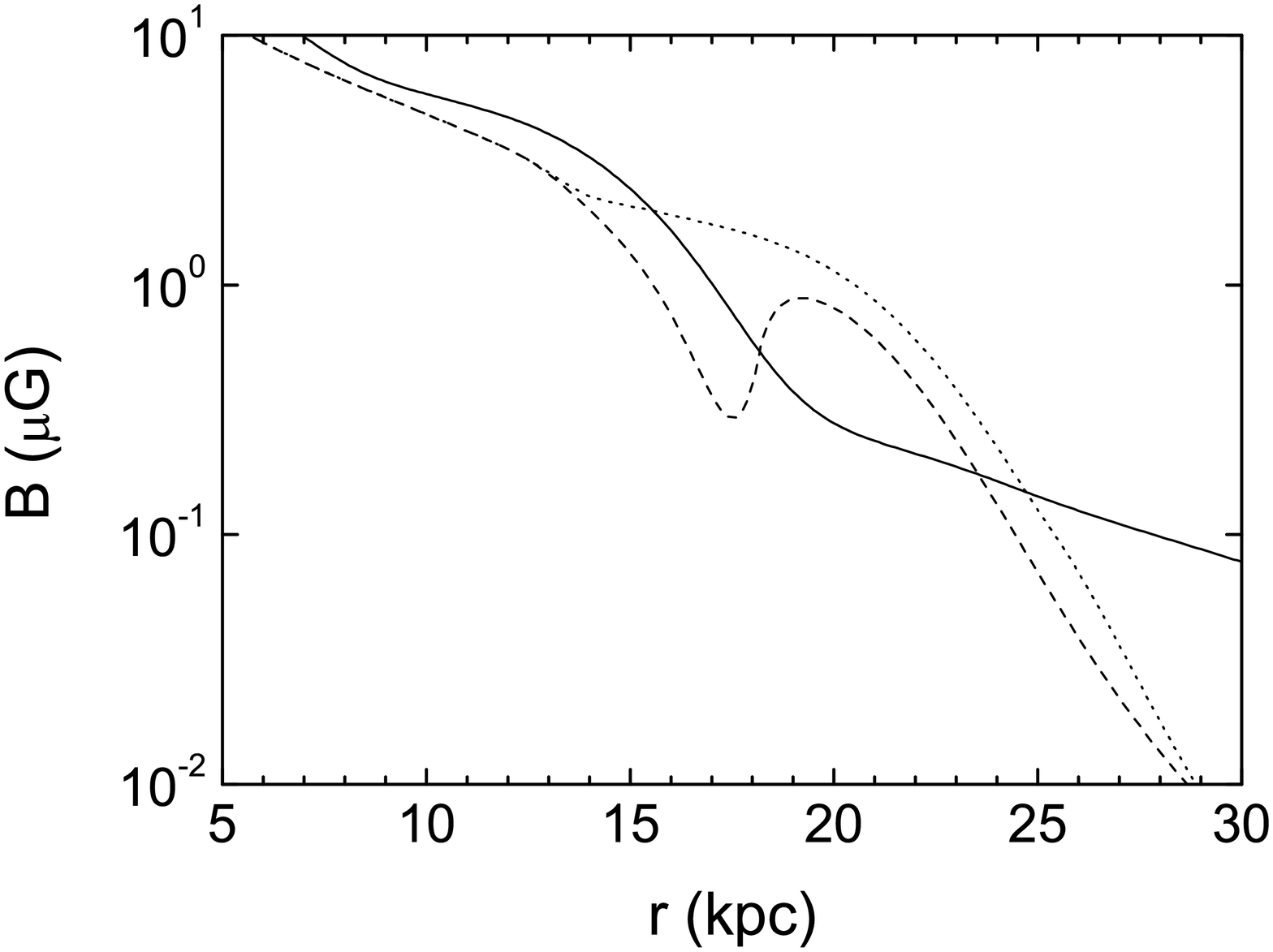}

\end{center}
\caption{Large-scale magnetic field strengths for different models
of the ionized gas component and initial conditions. The top panel
shows the strength of the final magnetic fields (i.e. at $10$ Gyr)
for different amplitudes of the initial conditions: curves from
bottom to top indicate $B_{0}=10^{-5} \mbox{ } \mu$G,
$B_{0}=10^{-4} \mbox{ } \mu$G, and $B_{0}=10^{-3} \mbox{ } \mu$G.
The middle panel shows the final field strengths for different
distances of the peak of the initial field from the centre: curves
from bottom to top show $r_{0}=6 \mbox{ kpc}$, $r_{0}=16 \mbox{
kpc}$, and $r_{0}=20 \mbox{ kpc}$. The bottom panel shows the
final magnetic field strengths for different models of the dynamo
scaleheight: the solid curve corresponds to model~1, the dashed
curve to model~2a, the dotted curve to model~2b.} \label{fig4}
\end{figure}

We assume an algebraic non-linear $\alpha$~quenching, $\alpha
\propto (1+(B/B^{*})^{-2})$, where $B^{*} \propto v \sqrt{4 \pi
\rho(r)}$ is the strength of the  field in equipartition with the
turbulent energy density field, which also depends on radius
($\rho$ is the interstellar gas volume density),
$B=\sqrt{B_{r}^{2}+B_{\varphi}^{2}}$. This non-linear
parametrization is quite empirical and implies that $\alpha$ is
reduced if the magnetic field strength is close to $B^{*}$. We
note that a more sophisticated dynamical quenching could be used
(Sur et al. 2007; Mikhailov 2013), but this is not necessary for
our purposes. The volume density is calculated by taking $\rho
\propto \frac{\Sigma_{\rm i}}{2 h_{\rm dyn}},$ where $\Sigma_{\rm
i}$ is as described in Sect.~3. For simplicity, we take $B^{*}
\approx 5 \mbox{ } \mu \mbox{G}$ at $r=8.5$\,kpc.

We parametrize the $\alpha$~effect as $\alpha \sim \Omega \,
l^{2}/h_{\rm dyn}$, where $h_{\rm dyn}$ is obtained from the
models for the ionized gas component.  There is some freedom in
choosing the model of the interstellar gas. In models with
constant dispersion of the turbulent velocity $v$, the scaleheight
$h_{\rm dyn}$ should increase in the outer parts; otherwise, with
$h_{\rm dyn}=\rm{const},$ $v$ decreases. Moreover, the model
allows choice of values for $v$ and $h_{\rm dyn}$, which are taken
to achieve a general agreement with the observational data
(Sect.~\ref{rad}). Model~1 uses constant $h_{\rm dyn}=1.0$\,kpc
and decreasing $v$. Models~2a and 2b use constant $v=10 \mbox{ km
s}^{-1}$ and $h_{\rm dyn}$ increasing in the outer parts of the
galaxy. We calculated the magnetic field strength for various
models of the ionized gas component and confirmed that the results
are quite similar for the different parameters.

For the initial conditions we take
$$B=B_{0}\left(\frac{r}{r_{0}}\right) \exp (-r/r_{0}),$$ where
$B_{0}$ takes different values (see Fig.~\ref{fig4}); using a
random seed field does not affect the results  significantly. Then
we performed our modelling for $0<r<50\mbox{ kpc}$, assuming that
$B(0)=B(50 \mbox{ kpc})=0$. We experimented with different ranges
of $r$, changing the outer limit from $30$ to $100$\,kpc, and the
results remained qualitatively similar. The magnetic fields are
still significant for radii as large as $r \approx 15 - 30 \mbox{
kpc}$. However, to exclude boundary effects near $r=r_{\rm max}$,
$r_{\rm max}$ needs to be substantially larger than these values.
We use $r_{\rm max}=50 \mbox{ kpc}$ as the standard value, which
seems to be quite satisfactory for our purposes.

Of course, our model could be developed to include various more
complicated effects, such as spiral arms (Moss et al. 2013) or
magnetic field reversals (e.g. Moss et al. 2012; Moss \& Sokoloff
2013), but here we consider only the simplest situation.

\section{Results}
\label{results}

We solved the dynamo equations (\ref{eqBr}) and (\ref{eqBphi}) for
our various assumptions about the shape of the disc of ionized gas
to obtain the evolution of the large-scale (regular) magnetic
field in the outer parts. Figure~\ref{fig2} describes the results
for the regions where magnetic fields have significant values (up
to $r=20 - 30 \mbox{ kpc}$). We halted the calculations at the
time 10~Gyr, close to the present time, and refer to the resulting
magnetic fields as the ``final'' ones. These and the other figures
in this paper show the strength of the  large-scale field,
calculated as $B=(B_{r}^{2}+B_{\varphi}^{2})^{1/2}$ (the no-$z$
model does not explicitly calculate the $z$~component, which is
small).  We stress that our models contain the azimuthal magnetic
field $B_\varphi,$ as well as the radial $B_r$; however, we note
that the computed magnetic field is very close to azimuthal
(Fig.~\ref{fig3}). The pitch angle in different parts of the
galaxy varies between $4^\circ$ and $15^\circ$. As expected, the
strength of the large-scale magnetic field is greater in the inner
parts of the galactic disc than in the remote regions. It is,
however, important that the magnetic field grows with time at the
disc periphery up to radii $15-30$\,kpc, becoming substantial (but
still weaker than in the inner regions) after $t\approx 10$\,Gyr.
In other words, the galactic dynamo can produce regular magnetic
fields of astronomically interesting strengths at galactocentric
radii of up to several tens of kpc. In any case, the large-scale
magnetic field at such radii is much larger than the lower limit
suggested by Neronov \& Semikoz (2009) for intergalactic space.

\begin{figure}
\begin{center}
\includegraphics[width=8cm]{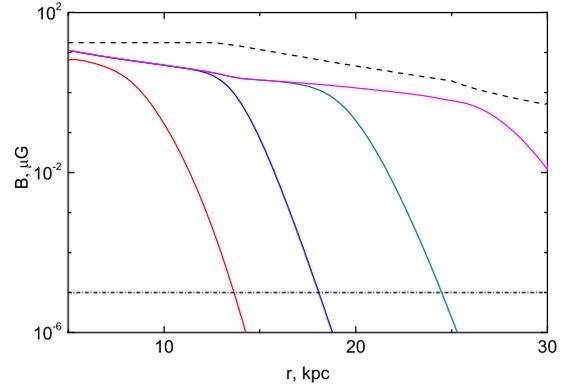}
\end{center}
\caption{Growth of an initial large-scale field concentrated in
the inner parts of the disc. Solid curves from bottom to top show
$t= 2,\ 5,\ 10,\ 20 \mbox{ Gyr}$; the dashed curve shows the
equipartition field strength; the dot-dashed line shows the field
strength for which the speed of propagation is estimated. See
text.} \label{fig5}
\end{figure}

\begin{figure}
\begin{center}
\includegraphics[width=8cm]{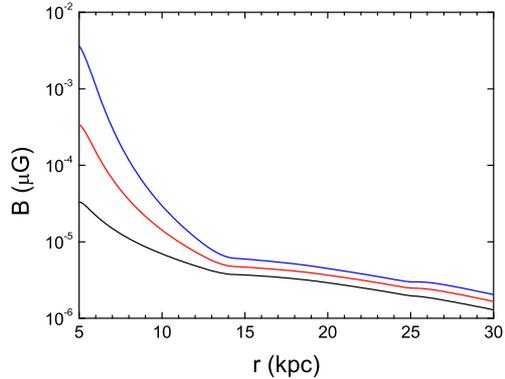}
\end{center}
\caption{Pure exponential growth of a weak magnetic field. The
curves from bottom to top show $t=0.2 \mbox{ Gyr}$, $t=0.4 \mbox{
Gyr,}$ and $t=0.6 \mbox{ Gyr}$.} \label{fig6}
\end{figure}

A comparison of magnetic field evolution for the various parameters describing the ionized gas distribution is illustrated in the panels of Fig.~\ref{fig2}. Although some details of the evolution are model dependent,  the
presence of substantial magnetic fields in the remote regions of galactic discs persists in all models.

In Fig.~\ref{fig2} (left-hand panels) we show the solutions for
$B(r)$ at different times. In the right-hand panels, we reproduce
$B(t)$ for different parts of the galaxy. A general conclusion is
that the large-scale magnetic field grows quite rapidly at
$r<20$\,kpc. Although it still grows at larger distances
($r>20$\,kpc), the growth rate is very slow there. We note that
Model~2a demonstrates a local minimum at $r \approx 17$\,kpc. This
is connected with the constant half thickness of the disc out to
$r = 18$\,kpc. The dynamo number is quite small near to
$r=18$\,kpc, and the field grows slowly. It looks implausible that
this feature is real,  and probably the disc becomes thicker at
smaller radii. For us it is important, however, that even for this
model, there is a quite noticeable magnetic field in the outer
parts of the galaxy.

Next, we demonstrate that our model is quite robust. The final
fields shown in Fig.~\ref{fig4} (top panel) are about $10 \mbox{ }
\mu \mbox{G}$ in the central parts of the galaxy and about
$10^{-1} \mbox{ } \mu \mbox{G}$ in the outer regions. The middle
panel of Fig.~\ref{fig4} shows the final fields for various radial
distances $r_{0}$ of the position of the maximum of the initial
field. As can be seen, the final magnetic field is quite similar
for the different models and does not depend
significantly on the initial field strength. The bottom
panel of Fig.~\ref{fig4} shows the final magnetic field for
various models of the partially ionized gas layer. The results do
not differ significantly, so we can claim that in all
cases the magnetic field in the outer parts is about $10^{-1}
\mbox{ } \mu \mbox{G}$. All these figures show the magnitude  of
the field as $B=(B_{r}^{2}+B_{\varphi}^{2})^{1/2}$.

\subsection{Propagation of the magnetic field to the remote parts of~the~disc}

A more detailed inspection of our model shows that there are at
least two possibilities for obtaining a significant magnetic field
in the remote parts of a galactic disc. Either the magnetic field
can be generated {\it in situ} near the disc periphery from a seed
field by local dynamo action or the magnetic field can be produced
in the central region of the galaxy and then transported to the
remote regions by the joint action of a dynamo and turbulent
diffusivity. We illustrate the second possibility, which follows
Moss et al. (1998).  The limited radial extent of standard
galactic dynamos has usually been insufficient to identify this
positive role of radial diffusion.

We consider a magnetic field that grows at a~rate~$\Gamma$. The
corresponding eigenfunctions are concentrated in the inner regions
of the disc and decay in the outer parts of the disc as $\exp
(-r/r_{\rm d})$ with $r_{\rm d}=(\eta / \Gamma)^{1/2}$. Then $B
\propto \exp (- r/r_{\rm d})$ and the radius at which $B$ exceeds
a given level propagates along $r$ to the remote parts of disc
with the speed

\begin{equation}
V_{\rm prop}\approx 2 \sqrt{\Gamma \eta}.
\label{KPP}
\end{equation}
This process of propagation is similar to the
Kolmogorov-Petrovsky-Piskunov effect (Kolmogorov 1937). We chose
an initial condition concentrated at the inner part of the disc
(Fig.~\ref{fig5}) and find that the point at which $B=
10^{-5}\,\mu$G moves with a speed   $V_{\rm prop} \approx  2
\mbox{ kpc Gyr}^{-1}$, which agrees with the prediction of
Eq.~(\ref{KPP}). For  a seed field that smoothly decays  with
radius (Fig.~\ref{fig6}), the shape of magnetic field distribution
remains more or less the same, while its amplitude grows with
time.

\section{Observational tests}
\label{tests}

Reliable data on the magnetic field distribution in the outer disc
regions, which can be compared with those theoretically
expected, may be obtained from radio observations of non-thermal
radiation of galaxies at large radial distances. Total radio
emission is stronger and hence easier to detect, but its detection
is limited by the background of unresolved sources
(``confusion''). Furthermore, inverse Compton losses of the cosmic
ray electrons interacting with photons of the cosmic microwave
background (CMB) makes it impossible to detect magnetic fields
with total strengths below $3.3\,(1+z)^2 \mbox{ } \mu
\mbox{G}$  \footnote{The local CMB energy density of $4.3 \,\,
10^{-13}$~erg~cm$^{-3}$ corresponds to that of a magnetic field of
strength $3.3\,\mu \mbox{G}$.}. Large-scale fields can also be
detected via the Faraday rotation of polarized background sources.
Since the rotation angle increases with the thermal electron
density, the line-of-sight component of the regular field, and the
square of the observation wavelength, low-frequency telescopes are
particularly well suited to measuring Faraday rotation in weak
fields. A regular field of $0.1 \mbox{ } \mu \mbox{G}$ with a
coherence length of 1\,kpc and an electron density of $10^{-3}
\mbox{ } \mbox{cm}^{-3}$ results in a Faraday rotation angle of
about 0.3\,rad at 150\,MHz (2\,m wavelength). Even regular fields
or electron densities that are lower by factors of 3 still
generate a measurable amount of Faraday rotation. However, the
main limitation comes from Faraday rotation in the Galactic
foreground that has to be subtracted (Oppermann et al. 2012). A
sufficiently large number of background sources is needed,
which restricts the observations to galaxies with a large
angular size on the sky.

The galaxies that possess extended HI discs (e.g. Sanders \&
Noordermeer 2007; Lelli et al. 2010) or UV discs (Thilker et al.
2007), which spread far beyond the conventionally defined optical
borders, are of special interest. Star formation currently taking
place in the outer UV discs guarantees the presence of
relativistic electrons and ionized gas. As the radio brightness is
expected to be low in the extended discs, modern low-frequency
arrays such as LOFAR,  MWA, and the planned SKA may be suitable
for determining the radii of the outer limits of magnetic fields
generated in differentially rotating galactic discs.

\begin{acknowledgements}
RB acknowledges support from the DFG Research Unit FOR1254. This
work was supported by Russian Foundation for Basic Research
(projects 12-02-00685). DS is grateful to MPIfR and Prof.
Michael~Kramer for hospitality during his stays in Bonn. EM
acknowledges support from the Dynasty Foundation. The authors
thank Marita Krause and the anonymous referee for useful
recommendations that helped us to improve the paper.
\end{acknowledgements}


\begin{thebibliography}{99}

\bibitem[Arshakian et~al. 2009]{Arshakian}
Arshakian, T.~G., Beck, R., Krause, M., \& Sokoloff, D. 2009, \aap,
  494, 21

\bibitem[Basu \& Roy 2013]{Basu2013}
Basu, A., \& Roy, S. 2013, \mnras,  433, 1675

\bibitem[Beck 2007]{Beck2007}
Beck, R., 2007, \aap, 470, 539

\bibitem[Beck 2013]{Beck2013}
Beck, R., \& Wielebinski, R. 2013, in: {\em Planets, Stars and Stellar Systems}, Vol.~5,
eds. T.~D.~Oswalt \& G.~Gilmore, Springer, Dordrecht, p.~641

\bibitem[Beck et al. 1996]{betal96}
Beck, R., Brandenburg, A., Moss, D., et~al. 1996, \araa, 34, 155

\bibitem[Berkhuijsen et~al. 2006]{Berkhuijsen06}
Berkhuijsen, E.~M., Mitra, D., \& M\"uller, P. 2006, Astr. Nachr., 327, 82

\bibitem[Berkhuijsen \& M\"uller 2008]{Berkhuijsen08}
Berkhuijsen, E.~M., \& M\"uller, P. 2008, \mnras, 390, L19

\bibitem[Bigiel et~al. 2010]{Bigiel10}
Bigiel, F., Leroy, A., Walter, F., et~al. 2010, \aj, 140, 1194

\bibitem[Bovy et~al. 2012]{Bovy12}
Bovy, J., Allende Prieto, C., Beers, T.~C., et~al. 2012, \apj, 759, 131

\bibitem[Brand \& Blitz 1993]{Brand93}
Brand, J., \& Blitz, L. 1993, \aap, 275, 67

\bibitem[Brandenburg 2014]{Brandenburg14}
Brandenburg, A. 2014,  [arXiv:1402.0212]

\bibitem[Case \& Bhattacharya 1998]{Case98}
Case, G.~L., \& Bhattacharya, D. 1998, \apj, 504, 761

\bibitem[Chamandy et al. 2013]{Chamandy13}
Chamandy, L., Subramanian, K., \& Shukurov, A. 2013, \mnras, 428, 3569

\bibitem[Chy{\.z}y \& Beck 2004]{Chyzy04}
Chy{\.z}y, K.~T., \& Beck, R. 2004, \aap, 417, 541

\bibitem[Cordes \& Lazio 2002]{Cordes02}
Cordes, J.~M., \& Lazio, T.~J.~W. 2002, [arXiv:astro-ph/0207156]

\bibitem[Cordes \& Lazio 2003]{Cordes03}
Cordes, J.~M., \& Lazio, T.~J.~W. 2003, [arXiv:astro-ph/0301598]

\bibitem[Cox 2005]{Cox05}
Cox, D.P. 2005, \araa, 43, 337

\bibitem[Do et~al. 2013]{Do2013}
Do, T., Martinez, G.~D., Yelda, S., et~al. 2013, \apjl, 779, L6

\bibitem[Donner \& Brandenburg 1990]{Donner90}
Donner,~K.~J., \& Brandenburg,~A. 1990, \aap, 240, 289

\bibitem[Ferri{\`e}re 2001]{Ferriere01}
Ferri{\`e}re, K.~M. 2001, Rev. Mod. Phys., 73, 1031

\bibitem[Fich et~al. 1989]{Fich89}
Fich, M., Blitz, L., \& Stark, A.~A. 1989, \apj, 342, 272

\bibitem[Fletcher et~al. 2011]{Fletcher11}
Fletcher, A., Beck, R., Shukurov, A., Berkhuijsen, E.~M., \&
  Horellou, C. 2011, \mnras, 412, 2396

\bibitem[Fraternali \& Tomassetti 2012]{Fraternali12}
Fraternali, F., \& Tomassetti, M. 2012, \mnras, 426, 2166

\bibitem[Gaensler et~al. 2008]{Gaensler08}
Gaensler, B.~M., Madsen, G.~J., Chatterjee, S., \& Mao, S.~A. 2008,
  \pasa, 25, 184

\bibitem[Gentile et~al. 2007]{Gentile07}
Gentile, G., Salucci, P., Klein, U., \& Granato, G.~L. 2007, \mnras,
  375, 199

\bibitem[Ghez et~al. 2008]{Ghez2008}
Ghez, A., Salim, S., Weinberg, N.~N., et~al. 2008, \apj, 689, 1044

\bibitem[Gillessen et~al. 2009]{Gillessen2009}
Gillessen, S., Eisenhauer, F., Fritz, T.~K., et~al. 2009, \apj, 707, L114

\bibitem[Gressel et~al. 2013]{G13}
Gressel, O., Elstner, D., \& Ziegler, U. 2013, \aap, 560, A93

\bibitem[Haffner et~al. 2009]{Haffner09}
Haffner, L.~M., Dettmar, R.-J., Beckman, J.~E., et~al. 2009, Rev. Mod. Phys., 81, 969

\bibitem[Holwerda et~al. 2012]{Holwerda12}
Holwerda, B.~W., Pirzkal, N., \& Heiner, J.~S. 2012, \mnras, 427, 3159

\bibitem[Hunter et~al. 2011]{Hunter11}
Hunter, D.~A., Elmegreen, B.~G., Oh, S.-H., et~al. 2011, \aj, 142, 121

\bibitem[Jalocha et~al. 2012]{Jalocha}
Ja{\l}ocha, J., Bratek, L., Peckala, J., \& Kutschera, M. 2012, \mnras, 427, 393

\bibitem[Kalberla \& Dedes 2008]{Kalberla08}
Kalberla, P.~M.~W., \& Dedes, L. 2008, \aap, 487, 951

\bibitem[Kalberla et~al. 2007]{Kalberla07}
Kalberla, P.~M.~W., Dedes, L., Kerp, J., \& Haud, U. 2007, \aap, 469, 511

\bibitem[Kasparova \& Zasov 2008]{Kasparova08}
Kasparova, A.~V., \& Zasov, A.~V. 2008, Astronomy Letters, 34, 152

\bibitem[Kolmogorov et~al. 1937]{Kolmogorov37}
Kolmogorov, A.~N., Petrovsky, I.~G., \& Piskunov, N.~S. 1937, Bull. Moscow State Univ., 1, 6

\bibitem[Lacki \& Beck  2013]{LB13}
Lacki, B.C., \& Beck, R. 2013, \mnras, 430, 3171

\bibitem[Lelli et al. 2010]{Lelli2010}
Lelli, F., Fraternali, F., \& Sancisi, R. 2010, \aap, 516, 11

\bibitem[Lewis \& Freeman (1989)]{Lewis89}
Lewis, J.~R., \& Freeman, K.~C. 1989, \aj, 97, 139

\bibitem[Lyne et~al. 1985]{Lyne85}
Lyne, A.~G., Manchester, R.~N., \& Taylor, J.~H. 1985, \mnras, 213, 613

\bibitem[Mao et~al. 2010]{Mao10}
Mao, S.~A., Gaensler, B.~M., Haverkorn, M., et~al. 2010, \apj, 714, 1170

\bibitem[Mera et~al. 1998]{Mera98}
Mera, D., Chabrier, G., \& Schaeffer, R. 1998, \aap, 330, 953

\bibitem[Mikhailov et~al. 2012]{Mikhailov}
Mikhailov, E.~A., Sokoloff, D.~D., \& Efremov, Y.~N. 2012, Astronomy Letters, 38, 611

\bibitem[Mikhailov 2013]{Mikhailov13}
Mikhailov, E.~A. 2013, Astronomy Letters, 39, 414

\bibitem[Moss 1995]{m95}
Moss, D. 1995, \mnras, 275, 191

\bibitem[Moss et al. 1998]{Moss98}
Moss, D., Shukurov, A., \& Sokoloff, D. 1998, GAFD, 89, 285

\bibitem[Moss \& Sokoloff 2011]{Moss1}
Moss, D., \& Sokoloff, D. 2011, AN, 332, 88

\bibitem[Moss \& Sokoloff 2013]{MS2013}
Moss, D., \& Sokoloff, D. 2013, GAFD, 107, 497

\bibitem[Moss et al. 2012]{our2012}
Moss, D., Stepanov, R., Arshakian, T.~G., Beck, R., Krause, M., Sokoloff, D. 2012, \aap, 537, A68

\bibitem[Moss et al. 2013]{our2013}
Moss, D., Beck, R., Sokoloff, D., Stepanov, R., Krause, M., Arshakian, T.~G. 2013, \aap, 556, A147

\bibitem[Moss et al. 1998]{metal98}
Moss, D., Shukurov, A., \& Sokoloff, D. 1998, GAFD, 89, 285

\bibitem[Narayan \& Jog 2002]{Narayan02}
Narayan, C.~A., \& Jog, C.~J. 2002, \aap, 394, 89

\bibitem[Narayan et~al. 2005]{Narayan05}
Narayan, C.~A., Saha, K., \& Jog, C.~J. 2005, \aap, 440, 523

\bibitem[Neronov \& Semikoz (2009)]{ns09}
Neronov, A., \& Semikoz, D. V. 2009, PRD, 80, 123012

\bibitem[Neronov \& Vovk (2009)]{nvovk10}
Neronov, A. \& Vovk, I. 2010, Science, 328, 73

\bibitem[Opperman et al. 2012]{oetal12}
Oppermann, N., Junklewitz, H., Robbers, G., et al. 2012, \aap, 542, A93

\bibitem[Phillips 2001]{Phillips}
Phillips, A. 2001, GAFD, 94, 135

\bibitem[Poezd et~al. 1993]{Poezd93}
Poezd, A., Shukurov, A., \& Sokoloff, D. 1993, \mnras, 264, 285

\bibitem[Ruiz-Granados et~al. 2012]{Ruiz-Granados12}
Ruiz-Granados, B., Battaner, E., Calvo, J., Florido, E., \&
  Rubi{\~n}o-Mart{\'{\i}}n, J.~A. 2012, \apjl, 755, L23

\bibitem[Ruzmaikin et al. 1985]{Ruzmaikin85}
Ruzmaikin, A.~A., Sokoloff, D.~D., \& Shukurov, A.~M. 1985, \aap, 148, 335

\bibitem[Ruzmaikin et~al. 1988]{Ruzmaikin}
Ruzmaikin, A.~A., Shukurov, A.~M., \& Sokoloff, D.~D. 1988,
{\em Magnetic Fields of Galaxies}, Kluwer, Dordrecht

\bibitem[Sanders \& Noordermeer 2007]{Sanders2007}
Sanders, R.~H., \& Noordermeer E. 2007, \mnras, 379, 702

\bibitem[Schnitzeler 2012]{Schnitzeler12}
Schnitzeler, D.~H.~F.~M. 2012, \mnras, 427, 664

\bibitem[Siejkowski et~al. 2014]{S14}
Siejkowski, H., Otmianowska-Mazur, K., Soida, M., Bomans, D.~J., \& Hanasz, M. 2014, \aap, 562, A136

\bibitem[Smith et~al. 2010]{Smith10}
Smith, B.~J., Giroux, M.~L., Struck, C., \& Hancock, M. 2010, \aj, 139, 1212

\bibitem[Subramanian \& Mestel 1993]{sm93}
Subramanian, K., \&  Mestel, L. 1993, \mnras, 265, 649

\bibitem[Sur et~al. 2007] {Sur} Sur, S., Shukurov, A., \& Subramanian, K. 2007, \mnras, 377, 874

\bibitem[Thilker et al. 2007]{Thilker2007}
Thilker, D. A., Bianchi, L., Meurer, G., et~al. 2007, \apjs, 173, 538

\bibitem[Vallee 1994]{Vallee94}
Vall{\'e}e, J.~P. 1994, \apj, 437, 179

\bibitem[Walter et~al. 2008]{THINGS}
Walter, F., Brinks, E., de Blok, W.~J.~G., et~al. 2008, \aj, 136, 2563


\end{thebibliography}
\end{document}